\documentclass[pdflatex,sn-mathphys-num]{sn-jnl}

\usepackage{float}
\usepackage{graphicx}%
\usepackage{multirow}%
\usepackage{amsmath,amssymb,amsfonts}%
\usepackage{amsthm}%
\usepackage{mathrsfs}%
\usepackage[title]{appendix}%
\usepackage{xcolor}%
\usepackage{textcomp}%
\usepackage{booktabs}%
\usepackage{algorithm}%
\usepackage{algorithmicx}%
\usepackage{algpseudocode}%
\usepackage{listings}%
\usepackage{longtable}%

\theoremstyle{thmstyleone}%
\theoremstyle{thmstyletwo}%

\theoremstyle{thmstylethree}%

\begin{document}

\title{Automated Video-EEG Analysis in Epilepsy Studies: Advances and Challenges}


\author[1]{Valerii A. Zuev}
\author[2]{Elena G. Salmagambetova}
\author[2]{Stepan N. Djakov}
\author[1]{Lev V. Utkin}
\affil[1]{\small Peter the Great St.Petersburg Polytechnic University \normalsize \\
	
	\texttt{valerii\_zuev@edu.spbstu.ru, lev.utkin@gmail.com}}
\affil[2]{\small Medical Center "XXI Century"\hspace{0pt}, St. Petersburg \normalsize }

\date{}
\maketitle

\section*{Abstract}

Video-electroencephalography (vEEG) monitoring is currently the gold standard in the diagnosis of epilepsy.
Manual analysis of vEEG recordings is time-consuming and requires automated tools for seizure detection.
Recent advances in machine learning have shown promise in real-time seizure detection and prediction using EEG and video data.
However, the diversity of seizure symptoms, markup ambiguities, and the limited availability of multimodal datasets hinder progress.
This paper reviews the latest developments in automated video-EEG analysis and discusses the integration of multimodal data.
We also propose a novel pipeline for treatment effect estimation from vEEG data using concept-based learning, offering a pathway for future research in this domain.




\maketitle

\section{Introduction}

The brain functions by sending electrical impulses from one neuron to another.
Electroencephalography (EEG) is a method that records the electrical activity generated by the combined activity of billions of neurons in the brain using electrodes placed on the scalp.
An abnormal EEG signal may indicate the presence of diseases, mainly epilepsy.
Video-EEG (vEEG) is a type of EEG with an added camera to monitor the patient's movements.

Epilepsy affects people of all sexes and ages worldwide, with an estimated global incidence rate of approximately 61 per 100,000 person-years \cite{beghi_epidemiologyEpilepsy_2019}, making it one of the most common neurological diseases along with Alzheimer's disease, dementia, and multiple sclerosis.
With appropriate treatment, about $ 70\% $ of patients can live seizure-free; therefore, accurate and timely diagnosis of epilepsy is very important.

The diagnosis of epilepsy is based on three main components:
\begin{enumerate}
	\item Clinical data
	\item EEG, vEEG.
	The so-called routine EEG examination, which lasts for approximately 30 minutes, is gradually being replaced in clinical practice by a more accurate long-term vEEG monitoring
	\item MRI scan
\end{enumerate}
However, examination results are sometimes difficult to interpret, so both under- and overdiagnosis of epilepsy remains a problem.
According to a 2019 study, 25-30\% of patients with previously diagnosed epilepsy who did not respond to initial drug therapy have no epilepsy \cite{amin_roleOfEegInErroneousDiagnosis_2019}.

Better insights are provided by a long-term video electroencephalogram (vEEG) \cite{zorgor2021comparisionRoutineVideoEeg} that is recorded over several hours or days.
Reading an entire electroencephalogram takes a significant amount of time, and computational tools to automatically detect abnormalities could help clinicians.
In addition, real-time seizure detection and early prediction using the Epilepsy Monitoring Unit (EMU) signal are important for timely assistance \cite{xu2024shorterLatencyRealTimeSeizureDetection}.
EEG electrodes can be placed on the scalp or implanted.
The latter method produces a cleaner signal, but the former is non-invasive; invasive methods are typically used in neurosurgery.
In this article, we will focus on scalp EEG.
Here, we give an overview of recent studies on EEG-based, video-based, and video-EEG-based approaches to seizure detection; to the best of our knowledge, as of January 2025, no review papers on combined video-EEG seizure detection have been published yet.
Moreover, we propose a novel, integrated framework for explainable automated treatment effect estimation using concept-based learning.

The paper is structured as follows.
Section \ref{section:backgnd} provides a brief recap of EEG and video-EEG in the context of epilepsy, as well as a short history of studies on seizure / epileptiform activity detection and sleep stage classification using EEG, video, and video-EEG.
Section \ref{section:recentAdvances} gives a more detailed overview of recent work in this area and then introduces a scheme of our proposed pipeline for treatment effect estimation.
In Section \ref{section:challenges}, we discuss potential challenges for machine learning in the field of seizure detection.

\section{Background} \label{section:backgnd}
	\subsection{Epilepsy, EEG and video-EEG monitoring}\label{section:backgnd:epilepsyEeg}

Epilepsy is a chronic brain disorder characterized by recurring seizures that occur as a result of excessive neural discharges and are accompanied by various motor, sensory, and mental dysfunctions.
In 2022, after 5 years of research, International League Against Epilepsy (ILAE) published a set of documents updating the definitions of epilepsy syndromes and their classification \cite{blinov2022ilaeClassification}.
The main criteria for differentiation is the age at which epileptic seizures are first detected: <2 years, 2-12 years, other; and a separate group of idiopathic generalized epilepsies.
In each group, the syndromes are additionally classified into \textit{generalized} (i.e. when abnormal activity occurs in both hemispheres), \textit{focal} (localized in a distinct part of the brain), and generalized/focal depending on the seizure type.
Each syndrome is described in a unified template which includes, among other characteristics, seizure types (mandatory, typical, occasional, and exclusionary), and EEG findings (background, interictal epileptiform discharges, ictal patterns, and provoking factors) \cite{wirrell2022ilaeClassification}.

Epileptic seizures, according to the 2017 ILAE guidelines \cite{fisher2017ilaeSeizureClassification}, are grouped by their onset into the following classes.

\begin{enumerate}
	\item Focal onset (epileptiform activity begins in a limited area of the brain, in one hemisphere), further categorized by awareness (aware / impaired awareness) and movements (motor / nonmotor).
	Focal to bilateral tonic-clonic seizures are described as a separate entity
	\item Generalized onset
	\begin{enumerate}
		\item Motor, including tonic-clonic and other motor seizures
		\item Nonmotor (absence), further classified into typical and atypical depending on EEG patterns and seizure characteristics
	\end{enumerate}
	\item Unknown/unclassified (due to inadequate information or inability to place the type in other categories)
\end{enumerate}

Motor seizures may include \textit{automatisms} (repetitive, purposeless movements such as lip smacking or hand rubbing), \textit{ clonic movements} (rhythmic jerking in specific muscles), \emph{tonic movements} (sustained stiffening of muscles), \textit{hyperkinetic activity} (agitated movements, such as thrashing or pedaling), or \textit{behavior arrest} (sudden cessation of motor activity).
Non-motor seizures lack prominent physical movements like jerking or stiffening and are manifested more in internal changes (cognitive, e.g. impaired thinking, language difficulties, or memory disruption; sensory, e.g. tingling, numbness, or pain in specific body parts; emotional, e.g. sudden feelings of fear, joy, anger, or sadness without an obvious trigger; autonomic, e.g. changes in heart rate, breathing, sweating, or gastrointestinal sensations) or brief pauses in activity.
Non-motor seizures can sometimes be subtle, making them harder to identify.

EEG of patients with epilepsy is characterized by the so-called \emph{epileptiform activity}, which occurs both during seizures and outside of them (interictal epileptiform discharges -- IEDs).
Among the epileptiform patterns are high-amplitude spikes, sharp waves, benign epileptiform discharges of childhood, spike-wave complexes, polyspikes, hypsarrhythmia and photoparoxysmal response.
Clinicians usually try to identify any epileptiform activity and then decide for each instance whether a seizure occurred or not; therefore, epileptiform discharges recognition is an even more relevant problem than seizure detection.
However, the latter is more difficult; there are fewer public datasets with labeled epileptiform activity than with seizures, and there are many more papers on seizure detection in recent years than on epileptiform discharges recognition.

Long-term video-EEG monitoring combines EEG and camera surveillance to capture and investigate seizures more reliably, as well as to distinguish between epileptiform discharges and artifacts such as normal limbs movements or eye blinks.

	\subsection{Automatic analysis of EEG and video-EEG}

There are two main purposes for epileptiform discharges and seizure detection.
The first is to reduce the doctor's time spent on analyzing long EEG recordings, and to make navigation through the recording easier by introducing an automatic markup.
The second is to provide timely assistance to the patient (currently specially trained Seizure Response dogs are used for the same purpose).

Attempts to recognize abnormal activity in EEG patterns have been made since the last century. Some threshold-based and wavelet-based algorithms, including those for spike detection, are implemented in clinical software and used in practice \cite{wilson2002spike}.
Recent advances in machine learning methods, particularly deep learning, and the addition of video modality to the electroencephalogram open the door to potential algorithms for detecting seizures and IEDs in real time or even in advance.
Since EEG monitoring equipment is expensive, cumbersome, and inconvenient for patients, alternative seizure detection methods have been developed, e.g. based on audio or body movement sensors; more details are provided in Section \ref{section:recentAdvances}.

Much effort has been devoted to sleep stage classification from EEG data \cite{sekkal_automaticSleepStageClassification_2022}.
Although not directly related to seizure and epileptiform activity detection, sleep stage classification could assist in labeling the electroencephalogram.
The purpose of automatic event detection is not to replace a human expert, but rather to help him or her navigate through the recordings, much like a table of contents helps navigate a book.

Often, symptoms of the disease are characteristic of a certain stage of sleep, e.g. sleep terrors often occur in stage 3 of non rapid eye movement (NREM) sleep \cite{pollakSleepEncyclopedia_2010}; epileptiform discharges usually predominate in stages 1 and 2 of NREM sleep.
A study in mice shows that more than half of epileptic seizures are preceded by rapid eye movement (REM) sleep stage \cite{ikoma_remEpilepsy_2023}.
Normal sharp-wave sleep patterns often resemble epileptiform activity.
Therefore, automatic labeling of sleep stages could reduce the time spent on EEG analysis by humans and make navigation easier.

	\section{Recent Advances} \label{section:recentAdvances}
	\subsection{EEG data analysis}

There are commercial tools on the market capable of real-time seizure and IED detection, with BESA, Encevis and Persyst being three of the most popular platforms \cite{koren2021commercialSeizureDetection}.
While the software is proprietary and algorithms are not disclosed, we can get a general understanding of some principles behind Persyst from a 2017 paper co-authored by its creators \cite{scheuer2017persyst}.
Neural network rules (NN-rules), composed of ensembles of individual neural networks (NNs), are used to describe events in the electroencephalogram.
Each NN-rule takes a segment of electroencephalogram as input and encapsulates an expert-defined concept, for example, "if the amplitude and sharpness of an event are significantly greater than that of background activity, then the probability that the event is a spike increases".

Recently, a number of studies have reported successful application of modern deep learning (DL) models for seizure detection, e.g. CNN2D+LSTM proved capable of real-time seizure detection \cite{lee2022realTimeSeizureComparison}.  
Some of them use stochastic modeling for EEG representation and feature selection \cite{furui_non-gaussianity_2021, furui_epileptic_2024}.
A 2022 review \cite{ren2022eegPerformanceEvalutation} compared existing models across different metrics such as seizure prediction horizon (SPH), i.e. the average time before the program warns of a seizure onset.

In 2024, a novel model incorporating 1D-CNN, Bi-LSTM and GRU layers  \cite{mallick2024eegSeizuresLSTM} achieved high performance in seizure/non-seizure EEG segment classification (99-100\% accuracy) on a popular open Bonn University dataset.

In another research \cite{mironov2024neurosoft} authors used augmentations to boost the performance of their model designed for absence seizure detection and incorporated this model into an EMU software, which allowed to automatically control the patient's level of consciousness during seizures in real time by asking pre-determined questions like "what is your name?".

There are efforts to harness a popular transformer architecture to process EEG data.
Examples include open-sourced EEGformer \cite{wan2023eegformer} and EEGFormer by Microsoft Research \cite{chen2024eegformer}.

Overall, as of February 2025, a significant number of articles on EEG-based epileptic seizure detection are published monthly; there are some recent papers on IED detection, although not so many as on seizure detection.
We performed a search on Google Scholar using Publish or Perish software \cite{callanga_publishOrPerish_2024}, using the phrase "epileptic seizure detection OR prediction" as title words.
Results for 2024 -- February 2025 are provided in the Appendix: table \ref{table:modelsFrameworks} lists new models and computational frameworks described in papers; table \ref{table:reviewPapers} contains a list of reviews; in table \ref{table:miscStudies}, there are other papers (neither review nor new model/framework studies).
Some results were filtered out, such as those not related to the original query or removed from the web; however, most entries were retained (197 in total).
In addition, we conducted a search using the phrase "epileptiform AND detection" to find articles on interictal epileptiform discharges (IED) and other epileptiform activity detection; we found 10 papers published in 2024 -- March 2025, and they are listed in Table \ref{table:epileptiform}.

There is currently no single state-of-the-art model for seizure detection and prediction using EEG, which is partly due to the diversity and complexity of existing performance metrics and seizure types, as discussed in detail in Section \ref{section:challenges}.

Numerous articles on sleep stage classification have been published in the past three years.
An et al. \cite{an_fusionMethodSleepStage_2021} proposed a multimodal fusion method.
Eldele et al. \cite{eldele_attentionBasedSleepStage_2021} used an attention mechanism in their model.
In 2023, Eldele et al. \cite{eldele_self-supervised_2023} adopted a self-supervised approach.
Multiple works used CNNs \cite{khalili_temporalCNNsleepStage_2021, yang_sleepStageMarkovCNN_2021, abbasi_ensembleCNNsleepStage_2022, goshtasbi_sleepStagefcn_2022, huang_senetSleepStage_2022, li_deepSleepStage_2022, zhou_singlechannelnet_2022}, while other use solutions such as CNN+GRU \cite{pei_hybridSleepStage_2022}, deep residual networks \cite{olesen_residualNetSleepStage_2021} or spatiotemporal graph convolutional networks \cite{jia_multiViewSpatioTemporalGCN_2021}.
You et al. \cite{you_fourierSleepStage_2022} use time, frequency, and fractional Fourier transform features as inputs to their neural network.
Fox et al. \cite{fox_foundationalSleep_2025} train a foundational Transformer using several open PSG datasets.
A recent review of deep learning-based sleep stage classification \cite{yue_deepLearningSleepStageReview_2024} provides more information on this topic.

	\subsection{Seizure detection and sleep stage classification from video}
Sometimes EEG monitoring is inconvenient or even impossible, so numerous studies focused on seizure detection from video stream. 
There are commercial proprietary tools for video-audio seizure recognition: Nelli \cite{armand2022nelli}, used in clinic, and SAMi \cite{atwood2021seizureDetectionDevices}, suitable for nocturnal seizure detection in a variety of diseases, including epilepsy.

Yang et al. \cite{yang2021videoSeizureLSTM} used CNN+LSTM for tonic-clonic seizure detection from video stream and achieved a mean sensitivity of 88\% and a mean specificity of 92\%.
Venkatesh et al. \cite{venkatesh_epileptic_2024} used MediaPipe Solutions Suite to track human poses.

Authors of VSViG (video-based seizure detection via skeleton-based spatiotemporal graph) \cite{xu2024vsvig} fine-tune a real-time pose estimation network on a clinical dataset.
Their model extracts patches around body joints, embeds them using a convolutional network and then feeds into a spatiotemporal graph model which outputs probabilities of seizure onset.
They achieve an approximately 6\% error rate and zero false detection rate, being able to detect a seizure 13 seconds in advance. 

Recent reviews \cite{ahmedt2024deepLearningSeizureVideoReview, ANDERSSON2024videoSeizureExploratoryStudy} acknowledge that seizure semiotics vary widely, and no current model is able to reliably detect all possible types.
One of the promising directions proposed in \cite{ahmedt2024deepLearningSeizureVideoReview} is the construction of a foundation model for processing epilepsy videos. \\

Several attempts have been made to design a model for sleep stage classification from video stream \cite{nochino_sleepStageCamera_2019, van_meulen_cameraBasedSleepStaging_2023, carter_sleepStageInfraredVideo_2023} in an effort to provide a more affordable and less obtrusive alternative to PSG.
Awais et al. \cite{awais_hybridDcnnSvmSleepStatesFromVideo_2021} designed a hybrid DCNN-SVM model for classifying facial expressions into sleep and non-sleep using EEG to create ground truth labels; perhaps a similar approach could be applied to sleep stage classification.
Carter et al. \cite{carter_sleepStageNearInfraredVideo_transformers_2024} designed SleepVST - a transformer model which is claimbed to be a current state-of-the-art in sleep staging from video.

	\subsection{Combined video-EEG analysis} \label{section:SeizureDetectionCombined}

Several works focused on video-based seizure detection assume that using other modalities such as audio and EEG may improve performance \cite{yang2021videoSeizureLSTM, ahmedt2024deepLearningSeizureVideoReview}.
Nevertheless, the simultaneous use of video and EEG remains an under-researched topic.
In 2017, Aghaei et al. \cite{aghaei2017videoEEG} extracted handcrafted features from video and EEG, then fed them into the K-nearest neighbors classifier.
Wu et al. \cite{wu2022videoEEGrolandicEpilepsy, cao2024synchronizedVideoEEG} presented a more advanced version of the same approach, achieving 99.19\% precision and 99.78\% recall for a classification task on 6-second segments.
However, this model is far from real-world application, since such metric values mean that the program will produce several false positives within one hour.
Lin et al. \cite{lin_vepinet_2024} used a similar pipeline to detect interictal epileptiform discharges. 

Yin et al.  \cite{yin2024multimodalEmotionRecognition} proposed a combination of transformer models for emotion recognition utilizing video, audio, and EEG modalities.
We suppose that such architecture could also be applied to seizure detection and sleep stage classification.

	\subsection{Other modalities}
Epileptic seizures can be detected by a combination of different sensors that record movements, pulse, and/or skin resistance.
Empatica \cite{regalia2019empaticaWristMonitors} produces wrist monitors with multiple sensors and proprietary seizure detection software.
There are multiple apps for smartphones (Epipal, SeizureTracker, SeizAlarm).
Some researchers attempt to detect seizures from audio signals, for example, Ashan et al. \cite{ahsan2019audio} trained a convolutional neural network to classify audio segments.
Fernandez-Martin et al. \cite{fernandez-martin_towardsEpileptiformMEG_2024} developed an ML algorithm for epileptiform discharges detection using magnetoencephalography (MEG); however, MEG, while sometimes providing a better picture of epileptiform activity than EEG, is not currently widely used clinically.
Visual memory tests based on eye tracking \cite{fu_recognition_2025} are used to assess memory deficiency associated with certain types of epilepsy, but to our knowledge, eye tracking has not been used for epileptiform discharges detection.


In sleep monitoring, the current gold standard is polysomnography (PSG), which includes EEG, electrooculogram (EOG), electromyogram (EMG), and sometimes other measurements such as blood oxygen saturation \cite{pollakSleepEncyclopedia_2010}.
In video-polysomnography (vPSG), a camera is added.
However, a typical sleep monitor used for PSG includes only 4-6 scalp EEG electrodes.

There are methods for automatic sleep stage classification from PSG data \cite{ji_mixsleepnetSleepStage_2024, ren_psgSleepStage_2025}; each signal carries information that is more relevant to different sleep stages  \cite{faust_sleepStageScoringReview_2019}, in particular, EEG is good at distinguishing between different stages of non-rapid eye movement (NREM) sleep, and EOG can help distinguish between REM and NREM.
With the spread of wrist fitness trackers and watches, several sleep monitoring algorithms using their sensors have been developed \cite{boe_sleepStageClassificationWearableSensors_2019, rentz_commercialWearableSleepMonitoring_2021, miller_sixWearableDevicesForSleepValidation_2022, schlegel_lessonsFromWearableDevicesChallenge_2024}, but as of 2024 they remain at least slightly less accurate than PSG-based methods \cite{Nelson_performanceWatchSleepVsPSG_2024, schyvens_wristBandsVsPSG_2024, robbins_wearableSleepAccuracy_2024, svensson_vouraVsPSG_2024}; moreover, wearable device manufacturers typically do not provide access to raw sensor data \cite{de_zambotti_wearableSleepResearch_2024}.
There are research-grade wearable sleep monitoring devices that are generally preferred over PSG for younger patients; this test is called \textit{actigraphy} \cite{ong_trackersActigraphy_2024, miner_headbandEegActigraphy_2025}.
Several apps have been developed that use smartphone microphones to detect sleep stage from breathing sounds, and some of them show good agreement with PSG-based sleep staging \cite{lee_accuracyWearablesNearablesAirables_2023}; the SleepScore app uses a patented sonar technology based on the phone's microphone and speaker.
Yu et al. \cite{yu_wifi-sleep_2021} used an industry-grade Wi-Fi transmitter-receiver system to track breathing patterns during sleep.
	
	\subsection{Available datasets}\label{section:recentAdvances:availableDatasets}

There are numerous datasets of EEG recordings, and many of them contain samples from patients with epilepsy, sometimes with designated seizure events or IEDs.
\begin{enumerate}[noitemsep]
	\item TUH EEG corpus \cite{obeid2016tuh} -- the largest publicly available collection of data, consisting of several sub-collections.
	Years of EEG recordings, including over 500 hours of annotated seizure data.
	Among other subcollections, includes TUSZ (a sub-dataset with designated epileptic seizures) and TUEV (a sub-dataset with designated epileptiform patterns).
	Access is available upon request
	\item The dataset from the University of Bonn (2001) \cite{andrzejak2001nonlinearDeterministicStructuresTimeSeries}, sometimes (e.g. in \cite{abdallah_deep_2025}) reffered to as the UCI Epileptic Seizure dataset, as a preprocessed and restructured version of this dataset was formerly hosted in the UCI data repository \cite{wu_epilepticUci_2017}.
	Five subjects, about 40 minutes of EEG per each dataset.
	Available for download from the University Hospital Bonn website
	\item Freiburg EEG database (2003) \cite{winterhalder_seizureFreiburgData_2003, maiwald_comparisonFreiburgData_2004} -- 582 hours of intracranial EEG (iEEG) recordings.
	No longer available for download as it is superseded by the European Epilepsy Database
	\item Flint Hills Scientific, L.L.C., Public ECoG Database (2008) \cite{frei_fhs_2008, duun-henriksen_channelSelection_flintHillsDatabase_2012}.
	1419h of EEG from 10 patients, 59 seizures.
	Electrocorticography (ECoG) is a speciific type of iEEG where electrodes are placed directly on the surface of the brain's cortex.
	As of February 2025, the website of the database is no longer maintained
	\item Children's Hospital Boston-Massachusetts Institute of Technology (CHB-MIT) Scalp EEG (2009) \cite{shoeb2009mlSeizureDetectionTreatment} -- 23 pediatric subjects, >900 hours of EEG recordings, 198 annotated seizures
	\item EEG + MRI data gathered at Warsaw Memorial Child Hospital (2010) \cite{zwolinski_openDatabaseEegMri_2010} -- 23 subjects with drug-resistant epilepsy.
	As of March 2025, not available online
	\item Bern-Barcelona EEG database (2012) \cite{andrzejak_nonrandomnessBernBarcelonaDatabase_2012}, 5 epilepsy patients. Publicly available on Kaggle
	\item EPILEPSIAE (a European Epilepsy Database) (2012) \cite{ihle_epilepsiaeDatabase_2012} is a part of the EU-funded project "EPILEPSIAE" aiming at the development of EEG-based seizure detection and prediction algorithms.
	300 epilepsy patients. 
	Available for purchase
	\item Karunya Institute of Technology and Scieinces database (2013) \cite{selvaraj_eegKarunyaDataset_2013} -- 175 abnormal EEG recordings
	\item American Epilepsy Society Seizure Prediction Challenge (AES-SPC, in some studies referred to as KEPC -- Kaggle Epilepsy Prediction Challenge or "UPenn and Mayo Clinic's Seizure Detection Challenge") (2014) \cite{brinkmann_crowdsourcingAesSpcData_2016, temko_detectionUPennChallenge_2015}.
	Five dogs, two patients; 628h of iEEG, 48 seizures.
	Publicly available on Kaggle
	\item Neurology \& Sleep Center at Haus Khas, New Delhi (NSC-ND, sometimes referred to as Haus Khas) (2016) (6 subjects) \cite{swami_eeg_2016, swami_novel_nscNdData_2016} -- available on ResearchGate
	\item Melbourne University AES/MathWorks/NIH Seizure Prediction Challenge (2016) \cite{kuhlmann_epilepsyecosystemorg_2018} -- discontinuous iEEG recordings. Publicly available on Kaggle
	\item SWEC-ETHZ iEEG Database (2018) \cite{burrello_one-shot-swecEthz_2018} -- 16 patients, 99 iEEG recordings. Publicly available
	\item Helsinki University EEG dataset, known also as Neonate (2019) \cite{stevenson2019neonate} -- 39 neonates, 342 seizure events
	\item Siena Scalp EEG database (2020) \cite{detti2020sienaScalpEegData} -- 14 adult patients, $ \approx $ 128 hours of EEG recordings, 47 annotated seizures
	\item A small dataset by Khati Renuka \cite{khati_epilepticData_2020} -- lacks sufficient description.
	Publicly available at Mendeley data
	\item Aarhus University Hospital dataset (2020) \cite{kural_criteriaIED_2020} -- 100 subjects, designated interictal epileptiform discrharges (IEDs).
	Publicly available
	\item Amerucan university of Beirut Medical Center scalp EEG dataset (2021) \cite{nasreddine_epilepticData_2021} -- 6 subjects.
	Publicly available at Mendeley Data
	\item S.Nijalingappa Medical College Bagalkot (SNMC) dataset (2022) \cite{deepa_b_preprocessedSnmcData_2022} -- 11 patients, 2 seizures. Available at IEEE Data Port
	\item Seoul National University Hospital (SNUH) (2023) \cite{shin_snuhDataset_2023} -- available per request
	\item Prince Hospital Khulna (PHK) (2024) \cite{das_epileptic_2024} -- 10 patients, 578s of seizure and 578s of non-seizure signals per patient.
	Processed data available on GitHub; raw data available per request
	\item VEpiSet (2024) \cite{lin_eegVEpiSet_2025} -- 84 subjects, 20 minutes per subject, designated IEDs
\end{enumerate}

Some of them, such as Neonate \cite{stevenson2019neonate}, were labeled independently by several experts, which is especially valuable since the correlation between different markups is usually low.
In 2023, Wong et al. \cite{wong2023eegSeizureDatasetsReview} gave a review on open EEG seizure datasets, however, some open datasets listed above are not mentioned in that paper.
	
There are not many publicly available seizure videos, presumably largely because this type of data is considered more sensitive than electroencephalogram.
The authors of VSViG \cite{xu2024vsvig} made their dataset public, blurring the patients' faces.
Miron et al. \cite{miron2024epilepticVideoYouTube} collect data from YouTube video hosting.

To our knowledge, there is no open dataset of epileptic seizures with both video and EEG modalities.
There is an open dataset of combined EEG and video recordings for emotion recognition -- EAV (EEG-Audio-Video for emotion recognition) \cite{lee2024eavData}, and another dataset of overnight EEG recordings combined with infrared videos for sleep analysis -- SSSVE \cite{han_sssve_dataset_2022, han_seeingSleepStage_2022}.
However, SSSVE does not contain original video sequences, only video features.


\subsection{Treatment effect estimation}

\textit{Treatment effect} (TE) studies aim to estimate the difference in a patient's condition if he or she takes a drug or other treatment and if he or she does not.
The patient's condition is usually represented as a continuous variable $y$, which might be, for example, blood sugar level or life expectancy from the current moment; \textit{average treatment effect} (ATE) is defined as the difference between its two expected values -- with treatment $y^{T=1}$ and without treatment $y^{T=0}$, and is designed to describe the potential benefit of taking the treatment.
When each patient is associated with some features, such as gender, age, and blood pressure, we can estimate the so-called \textit{conditional average treatment effect} (CATE) which is defined as the ATE conditional on the patient's features.
A central problem in TE studies is observability, i.e. a patient cannot be treated and untreated at the same time, so it is not possible to build a machine learning model for CATE estimation directly from clinical data since we cannot observe both $y^{T=0}$ and $y^{T=1}$.
As a result, researchers use augmentations \cite{utkin2020teEstimationAugmentation} and synthetic data.

Treatment effect estimations have been proposed for a variety of settings, including TE for censored data \cite{kirpichenko2024benk, konstantinov2024survival_interpretableTrajectories}. Many works focus on TE estimation from tabular data with relatively low-dimensional features \cite{kunzel2019metalearners}, but sometimes inputs are more complex, e.g. include an electroencephalogram.
Another possible way to use EEG in TE studies is to derive $y$ from electroencephalogram or replace $y$ with some features derived from it, which may require constructing a modified version of CATE.

\begin{figure}[H]
	\centering
	\includegraphics[width=\textwidth]{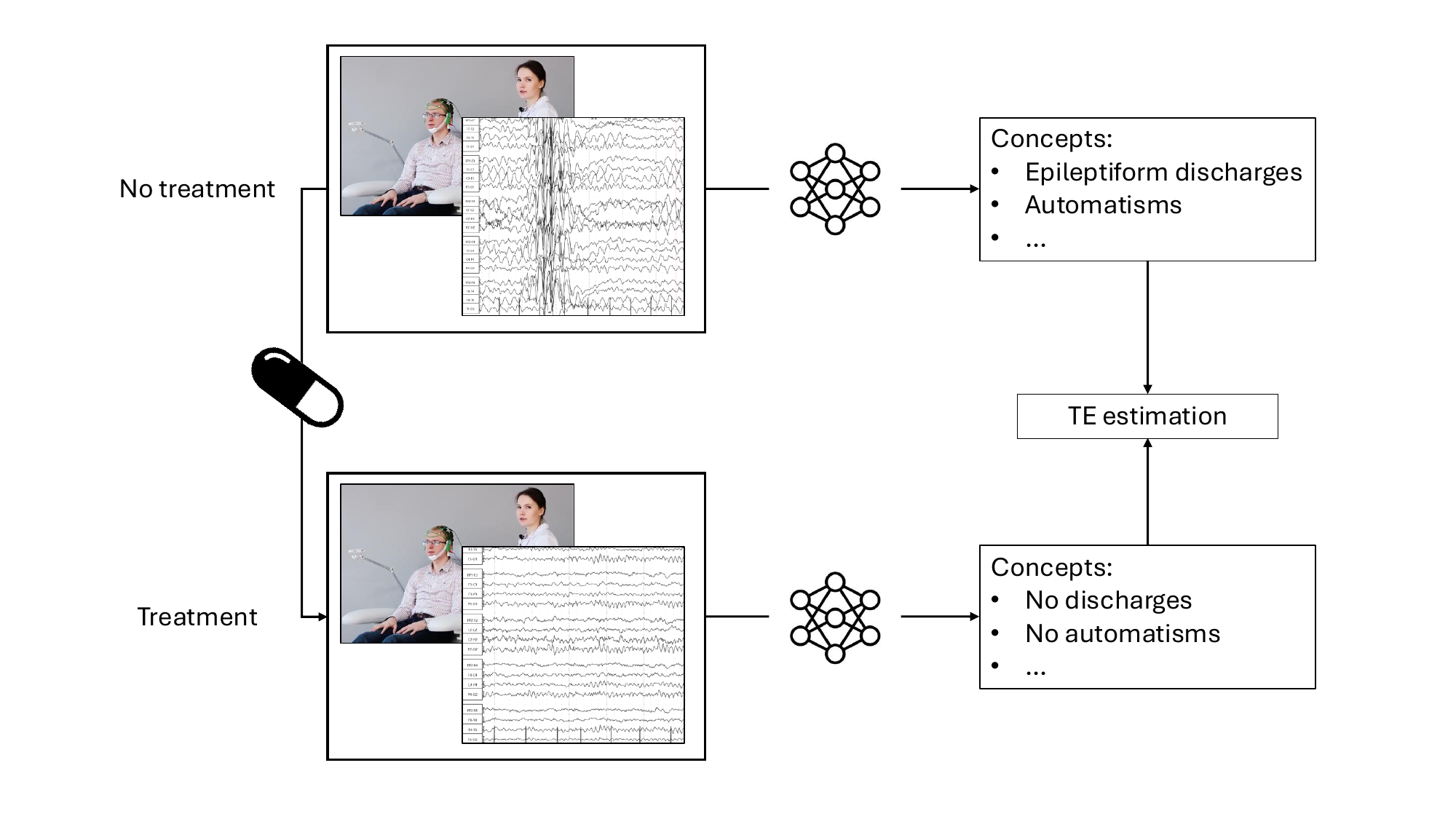}
	\caption{Proposed pipeline for treatment effect estimation}
	\label{fig:te_schema}
\end{figure}

Estimating TE from video-EEG data remains an understudied topic.
There are many works on both statistical \cite{ma2010statisticalEegTreatmentEffect} and machine learning-based \cite{watts2022eegTreatmentResponseMachineLearning} methods for treatment response prediction from EEG data.
Video-based TE estimation is described in \cite{ojanen2022videoTreatmentEffect}.
To the best of our knowledge, no studies have been conducted to date on the possibility of using combined video and EEG data for TE estimation.
We believe this is a promising direction for future research.
As we expect, it might be beneficial to apply concept-based learning \cite{poeta2023conceptBasedExplainableAi} to first infer human-understandable concepts from video-EEG data and then estimate TE from them (Fig. \ref{fig:te_schema}).
Concept-based Explainable Artificial Intelligence has been applied in many medical studies, including CT image processing \cite{dumaev2024conceptBasedExplainableMalignancyScoring}, with the aim of providing more transparent results than traditional ML systems and incorporating expert rules into automatic reasoning and decision making \cite{konstantinov2024expertRulesNNs, utkin2024fi_cbl}.
Several papers on concept-based EEG interpretation have been published recently \cite{knispel2022interpretableEEG, madsen2023conceptBasedEEGtransformer, brenner2024conceptBasedEEG}.
Pradeepkumar et al. \cite{pradeepkumar_interpretableSleepStageClassification_2024} proposed using cross-modal transformers for polysomnography (PSG)-based sleep stage classification using two modalities, EEG and electrooculogram (EOG).
Research using concepts for video processing is limited; and, although an automatic framework for concept extraction was proposed in 2022 \cite{jeyakumar2022automaticConceptExtractionVideo}, the corresponding study does not specifically focus on biomedical applications.

To better illustrate the proposed scheme, consider, for example, the rapid eye movement (REM) sleep stage \cite{pollakSleepEncyclopedia_2010} of a healthy individual.
It could be described using the following concepts:
\begin{enumerate}
	\item EEG:
	\begin{itemize}[noitemsep]
		\item No sleep spindles or K-complexes
		\item Low voltage (relatively to other stages) mixed frequency activity
		\item fewer vertex sharp transients than in stage 1 of NREM sleep
		\item (sometimes) distinctive "saw tooth" waves
	\end{itemize}
	\item Video:
	\begin{itemize}[noitemsep]
		\item Episodic, rapid eye movements (simultaneous to mixed frequency activity)
		\item Irregular contractions of the respiratory muscles
		\item Loss of skeletal muscle tone (REM atonia)
		\item Phasic jerks and twitches in all striated muscle
	\end{itemize}
\end{enumerate}

Description using such concepts can be used for sleep stage classification.
When solving the problem of epileptiform patterns detection, properties of a focal motor onset epileptic seizure \cite{anwar_epilepticSeizures_2020} could be described with the following concepts:
\begin{enumerate}
	\item EEG: localized epileptiform patterns (e.g. spikes, spike-waves, etc.)
	\item Video:
	\begin{itemize}[noitemsep]
		\item Jerking movements in the foot, face, or arm on the opposite side of the body, pedaling, etc.
		\item Lip-smacking or chewing
	\end{itemize}
\end{enumerate}

	\section{Challenges}\label{section:challenges}
	\subsection{Variety of symptoms}
As described in Section \ref{section:backgnd:epilepsyEeg}, seizures can be accompanied by a variety of symptoms ranging from loss of movement to rhythmic jerking of the limbs and various patterns of EEG activity.
This not only complicates seizure detection, but also increases the amount of data necessary for adequate training and benchmarking.
Absence seizures and other types of seizures characterized by sudden arrest of movements are not easy to detect in a video stream; we believe that combining different modalities will lead to better results. 
In addition, synthetic data might help in pre-training \cite{xu2022multichannelSynthetic}.

Another problem is that typical EEG and behavioral patterns vary significantly depending on age, gender, and sleep habits.
This makes seizure detection and sleep stage classification even more difficult \cite{wei_influenceSexAgeSeizureDetection_2024, moris_ageAffectsSleepStageClassification_2024}.
EEG can be noisy when measured on a hairy scalp; efforts are underway to develop electrodes that produce less noise, such as the 3D-printed hair-like electrodes described in a recent paper by Ahmed et al. \cite{ahmed_stick-and-playElectrodes_2025}.

It is important to distinguish between epilepsy and psychogenic nonepileptic seizures.
A recent study \cite{kucikiene_eegMicrostatesEpilepsyPsychogenicSeizures_epilepsia2024} shows that EEG microstates differ in these two types of seizures.
Another study \cite{ito_differentiatingEpilepsyNonEpilepsyGlut1_2024} identified typical clinical factors that clearly differentiate non-epileptic seizures from epileptic in Glut1 deficiency syndrome.
Fussner et al. \cite{fussner_epilepticPsychogenicDifferCnn_2024} utilized CNN to classify psychogenic and epileptic seizures, converting EEG signal to 2D images.
	
	\subsection{Markup ambiguities}

The markup in epilepsy datasets usually consists of intervals containing seizures. However, for diagnosis, not only seizures are important, but other features such as slow alpha rhythm.
The terms "epileptiform activity" and "epileptic seizure" are very complex and ambiguously defined (as mentioned in section \ref{section:backgnd:epilepsyEeg}), e.g. the ictal period often has no clear end; even an experienced clinician cannot sometimes reliably distinguish between epileptiform activity and events like hypnagogic hypersynchrony.  
Studies show low agreement between markups made by different experts \cite{mironov2024neurosoft, stevenson2019neonate}, so the labels are unreliable.
This led to an approach with multiple experts annotating the same dataset and then putting more weight during training to intervals with higher agreement.
Moreover, this once again highlights the complexity of the decision which is made by the expert not only based on a short period of recording around the current time frame, as almost all the machine learning models do, but also considering other factors such as the patient's age, diagnoses, their sleep cycle traits, etc.
Shama et al. \cite{shama_uncertainty-aware_2025} propose an uncertainty-aware Bayesian framework to handle label noise.

Inter-human agreement in sleep staging is also low \cite{van_gorp_certaintyAboutUncertaintySleepStaging_2022}, raising questions such as whether a model with higher agreement with human labeling is better in real-world applications.

	\subsection{Data availability} \label{section:challengesDataAvailability}

There are a significant number of open EEG recordings -- with epileptic seizures (both nocturnal and daytime), seizure-free (suitable for training a sleep stage classifier); sometimes there are labeled IEDs.
But the situation is different with videos, which hinders models training and comparison.
There is no systematic and standardized video seizure dataset other than \cite{xu2024vsvig}.
A review \cite{ahmedt2024deepLearningSeizureVideoReview} contains a table of video-based approaches for seizure detection with corresponding datasets; of over 30 items, most are private, one is composed of YouTube videos, and one is available as processed features, not the raw or even blurred video stream.
This is largely due to privacy concerns.
The same review \cite{ahmedt2024deepLearningSeizureVideoReview} suggests generating artificial data as a possible solution.
As mentioned above, synthetic data is also used to improve performance in the EEG modality \cite{xu2022multichannelSynthetic}.
Wang et al. \cite{wang_canine_2025} align human and canine EEG to boost their model's performance; perhaps a similar approach could be applied to videos.
Datasets with overnight video-EEG recordings of healthy individuals, such as SSSVE \cite{han_sssve_dataset_2022}, might be used for model pre-training.
However, as mentioned above in \ref{section:recentAdvances:availableDatasets}, the authors of SSSVE only made the video features publicly available, not the original videos.

	\subsection{Multi-modal data}

As stated above (Section \ref{section:SeizureDetectionCombined}), there is little work on epileptiform activity detection based on both video and EEG data.
In practice, when examining a patient, a doctor takes into account the data from video-EEG monitoring, the clinical condition, medical history, prescribed treatment, the MRI image and other factors such as DNA markers.
Goel et al. \cite{goel_utility_2024} use MRI data to improve EEG-based epileptiform activity detection, but only in focal epilepsy.
Dharani et al. \cite{dharani_integrative_2024} fuse EEG and MRI features, but their approach provides only a small improvement over EEG-based models.
This means that there is still much room for new methods combining different sources for better EEG annotation.

The authors suggest that one promising direction could be the creation of a multiparameter model of brain functioning, the parameters of which are set based on what is known about the patient: age, blood sugar, MRI image, typical patterns of EEG activity, PSG, etc.
Machine learning methods in a sense build an implicit model of the brain, but an explicit model would make it easier to incorporate multimodal data and expert knowledge, as well as improve the interpretability of the output.
This could also help generate synthetic data for pre-training ML models, which may be a next step in the approach described in \cite{xu2022multichannelSynthetic}.
Du et al. \cite{du_hyperdimensional_2024} harnessed hyperdimensional computing, which is a brain-inspired framework motivated by the observation that the cerebral cortex operates on high-dimensional data.
Zhang et al. \cite{zhang_efficient_2024} used a spiking neural network (SNN), which is a closer mimic of the real brain than a conventional deep neural network, to learn patterns from EEG; Gallou et al. \cite{gallou_online_2024} developed a specialized hardware implementation of an SNN. However, SNNs are relatively difficult to train, and it is unclear whether a relatively small number of artificial spiking neurons is a good model of the brain.
In our view, approaches to modeling brain activity on a larger scale using dynamical systems theory, such as neural mass models (NMMs) and neural field models (NFMs) \cite{breakspear_dynamicModelsBrainActivity_2017} offer a more promising direction;
recently, attempts have been made to combine such models with deep learning methods \cite{liu_forecasting_2024}.

Another possible direction is multimodal foundational models \cite{li2024multimodalFoundationalModelsSurvey} jointly trained on video and EEG data. However, the primary expected limitation of this approach is the lack of data described in Section \ref{section:challengesDataAvailability}.
Augmented and synthetic data might lead to improvements \cite{mironov2024neurosoft}, however, a good data synthesis or augmentation strategy will likely again require an underlying model of the process.

For sleep stage analysis using infrared videos, Han et al. used contrastive learning to align video to EEG modality \cite{han_cross-modal_2022, han_seeingSleepStage_2022}.
A similar approach might be applied to seizure detection.

\section{Conclusion}

Automated video-EEG analysis has significant potential to improve epilepsy diagnosis and treatment.
Advances in deep learning and multimodal data integration have enhanced real-time seizure and interictal epileptiform discharges detection and prediction capabilities.
However, challenges such as symptom variability, markup inconsistencies, and lack of publicly available data remain.
Future research should focus on developing robust multimodal models, leveraging synthetic data, and creating standardized datasets to improve model performance and generalizability.

The proposed concept-based learning pipeline for treatment effect estimation offers a promising direction for integrating expert knowledge and multimodal data, advancing personalized epilepsy care.

\newpage
\begin{appendices}

\section{Relevant papers published in 2024 -- January 2025}

\begin{center}
\begin{longtable}{p{7.5cm}|p{4cm}|p{2cm}}
\caption{Models and frameworks for EEG-based seizure detection (2024 - Jan 2025) } \label{table:modelsFrameworks} \\
	\toprule
	Paper Title (\textdagger  -- open-source code available) & Methodology  & Datasets \\
	\midrule\midrule
	Adadelta-CSA: Adadelta-Chameleon Swarm Algorithm for EEG-Based Epileptic Seizure Detection \cite{salini_adadelta-csa_2025}  & DNN with an efficient optimization algorithm & CHB-MIT \\
	\hline
	Uncertainty-Aware Bayesian Deep Learning with Noisy Training Labels for Epileptic Seizure Detection \cite{shama_uncertainty-aware_2025} \textdagger & Bayesian framework for EEG segment labeling & TUH \\
	\hline
	Novel EEG feature selection based on hellinger distance for epileptic seizure detection \cite{sadiq_novel_2025} & Particle Swarm Optimization (PSO) for feature selection & Bonn \\
	\hline
	Dual-Modality Transformer with Time Series Imaging for Robust Epileptic Seizure Prediction \cite{qin_dual-modality_2025} \textdagger & dual-modality NN (visual patterns from EEG + raw EEG) & CHB-MIT, Bonn \\
	\hline
	Evaluating Different Hybrid Learning Algorithms Based Grid Search Algorithm for Epileptic Seizure Zone Detection \cite{ozer_evaluating_2025} & mix of ML and statistical approaches & Bonn \\
	\hline
	PaFESD: Patterns Augmented by Features Epileptic Seizure Detection \cite{munoz_pafesd_2025} \textdagger  & Pattern matching (Dynamic Time Warping) & CHB-MIT \\
	\hline
	Patient-independent epileptic seizure detection using weighted visibility graph features and wavelet decomposition \cite{mohammadpoory_patient-independent_2025} & weighted visibility graph & CHB-MIT \\
	\hline
	Epileptic seizure detection in EEG signals via an enhanced hybrid CNN with an integrated attention mechanism \cite{mekruksavanich_epileptic_2025} & CNN + BiGRU + convolutional block attention & Bonn \\
	\hline
	Phase spectrogram of EEG from S-transform Enhances epileptic seizure detection \cite{liu_phase_2025} & Stockwell transform for feature extraction & CHB-MIT, Bonn \\
	\hline
	Trust EEG epileptic seizure detection via evidential multi-view learning \cite{liu_trust_2025} \textdagger & several NNs for high-level features & CHB-MIT \\
	\hline
	Epileptic Seizure Detection in SEEG Signals Using a Signal Embedding Temporal-Spatial–Spectral Transformer Model \cite{li_epilepticSpatialSpectralTransformer_2025} & multi-scale temporal-spatial-spectral transformer & Custom (XJSZ), Bonn, CHB-MIT \\
	\hline
	A rhythmic encoding approach based on EEG time-frequency image for epileptic seizure detection \cite{li_rhythmic_2025} & Reassigned Smoothed Pseudo Winger-Ville Distribution (alternative to wavelets) & Karunya \\
	\hline
	Diminished Mobilenet: A Lightweight Architecture for Epileptic Seizure Prediction Using Single-Channel Eeg \cite{jang_diminished_2025} & STFT + modified MobileNet & CHB-MIT, SNUH\\
	\hline
	Automated Epileptic Seizure Detection of EEG Signals Using Machine Learning \cite{jain_automated_2025} & SVM & CHB-MIT \\
	\hline
	Patient-Independent Epileptic Seizure Detection with Reduced EEG Channels and Deep Recurrent Neural Networks \cite{el-dajani_patient-independent_2025} & CNN-LSTM & Bonn \\
	\hline
	A deep learning paradigm with residual networks for enhanced epileptic seizure prediction \cite{chalil_deep_2025} & ResNet & CHB-MIT \\
	\hline
	EEG-Based Epileptic Seizure Prediction Using Variants of the Long Short Term Memory Algorithm \cite{akshita_eeg-based_2025} & LSTM variants & CHB-MIT \\
	\hline
	Deep Clustering for Epileptic Seizure Detection \cite{abdallah_deep_2025} \textdagger  & autoencoder + SVD + GMM & Bonn \\
	\hline
	Epileptic seizure prediction via multidimensional transformer and recurrent neural network fusion \cite{zhu_epilepticSeizureEegMultidimTransformer_2024} & STFT + transformer + LSTM + GRU & Bonn, CHB-MIT \\
	\hline
	An efficient channel recurrent Criss-cross attention network for epileptic seizure prediction \cite{zhu_efficient_2024} & STFT + CNN + channel attention + cris-cross attention & CHB-MIT \\
	\hline
	Residual and bidirectional LSTM for epileptic seizure detection \cite{zhao_bidirectLstmEegSeizure_frontCompNeurosc2024} & ResNet + BiLSTM & Bonn, TUH \\
	\hline
	Efficient and generalizable cross-patient epileptic seizure detection through a spiking neural network \cite{zhang_efficient_2024} & spiking NN (closely mimicking real brain neurons) & CHB-MIT, custom (PKU1st) \\
	\hline
	Cross-patient automatic epileptic seizure detection using patient-adversarial neural networks with spatio-temporal EEG augmentation \cite{zhang_cross-patient_2024} & data augmentation + NN with good cross-patient generalization & CHB-MIT \\
	\hline
	Epileptic Seisuze Detection Based on Channel Attention Mechanism \cite{zhang_epileptic_2024} & custom CNN-based architecture & CHB-MIT \\
	\hline
	A scheme combining feature fusion and hybrid deep learning models for epileptic seizure detection and prediction \cite{zhang_featureFusionEegSeizure_nat2024} & DWT + feature fusion + CNN-GRU attention & CHB-MIT \\
	\hline
	ConceFT-based epileptic seizure detection via transfer learning \cite{yousif_conceft-based_2024} & ConceFT (alternative to wavelet transform) & Bonn \\
	\hline
	Efficient Epileptic Seizure Detection Method Based on EEG Images: The Reduced Descriptor Patterns \cite{yahia_efficient_2024} & STFT + novel feature extraction from spectrogram images & CHB-MIT \\
	\hline
	Advanced Epileptic Seizure Detection Using Deep Learning and Bayesian Optimization \cite{yadav_advanced_2024} & VGG-16 + Double LSTM & Bonn \\
	\hline
	Shorter latency of real-time epileptic seizure detection via probabilistic prediction \cite{xu2024shorterLatencyRealTimeSeizureDetection} & Multi-scale STFT + 3D-CNN, claimed new SoTA on CHB-MIT & CHB-MIT, SWEC-ETHZ \\
	\hline
	Dynamic Functional Connectivity Neural Network for Epileptic Seizure Prediction Using Multi-Channel EEG Signal \cite{xu_dynamic_2024} & Dynamic Graph CN + CNN & CHB-MIT \\
	\hline
	Hybrid LSTM–Transformer Model for the Prediction of Epileptic Seizure Using Scalp EEG \cite{xia_hybrid_2024} & STFT + LSTM + Transformer & CHB-MIT \\
	\hline
	Enhancing Epileptic Seizure Detection with Random Input Selection in Graph-Wave Networks \cite{wu_enhancing_2024} & FFT + Graph WaveNet & TUSZ (part of TUH) \\
	\hline
	Combination of Channel Reordering Strategy and Dual CNN-LSTM for Epileptic Seizure Prediction Using Three iEEG Datasets \cite{wang_combination_2024} & multi-channel CNN-LSTM + importance-based channel reordering & Freiburg, SWEC-ETHZ, AES-SPC \\
	\hline
	Channel-Selection-Based Temporal Convolutional Network for Patient-Specific Epileptic Seizure Detection \cite{wang_channel-selection-based_2024} & Fisher score-based channel selection + custom NN & CHB-MIT, Siena \\
	\hline
	Automatic epileptic seizure detection using SVM techniques with EEG signals \cite{vidya_automatic_2024} & SVM & Bonn \\
	\hline
	The Deep Learning Based Epileptic Seizure Detection Using 2-layer Convolutional Network with Long Short-Term Memory \cite{vaithilingam_deep_2024} & CNN + LSTM & CHB-MIT, Bonn \\
	\hline
	Sparse wavelet representation of interictal epileptic activity for postsurgical seizure outcome prediction \cite{ulrich_sparse_2024} \textdagger & predefined wavelets for high-frequency oscillations detection & custom iEEG data \\
	\hline
	Automatic detection of epileptic seizure based on one dimensional cascaded convolutional autoencoder with adaptive window-thresholding \cite{timothy_aboyeji_automatic_2024} & unsupervised learning + 1D-cascaded convolutional autoencoder & CHB-MIT \\
	\hline
	Prediction of epileptic seizure based on image information derived from focal electrodes and deep classifiers \cite{teixeira_prediction_2024} & multiple strategies & EPILEPSIAE \\
	\hline
	Epileptic Seizure Detection Based on Path Signature and Bi-LSTM Network With Attention Mechanism \cite{tang_epilepticBiLstm_2024} & path signature feature extraction +Bi-LSTM & CHB-MIT, TUEP (part of TUH) \\
	\hline
	Epileptic Seizure Detection in Neonatal EEG Using a Multi-Band Graph Neural Network Model \cite{tang_epileptic_2024} & multi-band GNN & Neonate \\
	\hline
	Long-short term memory autoencoder using delta with beta bands of EEG enables highly accurate prediction of seizure outcome in Infantile Epileptic Spasms Syndrome of unknown etiology \cite{suzui_long-short_2024} & burst detection algorithm + LSTM-AE & custom \\
	\hline
	Federated Machine Learning for Epileptic Seizure Detection using EEG \cite{suryakala_federated_2024} & Graph Convolutional Nevwork + Federated Learning  & CHB-MIT, Bonn \\
	\hline
	Deep Learning for Epileptic Seizure Detection Using a Causal-Spatio-Temporal Model Based on Transfer Entropy \cite{sun_deep_2024} & Graph Attention Network + BiLSTM & SWEC-ETHZ, custom \\
	\hline
	Smart Societal Optimization-based Deep Learning Convolutional Neural Network Model for Epileptic Seizure Prediction \cite{sonawane_smart_2024} & CNN + intelligent societal optimization & CHB-MIT \\
	\hline
	EEG Data Classification by Hybrid Deep Learning for Epileptic Seizure Prediction \cite{slama_eeg_2024} &1D-CNN + BiLSTM & CHB-MIT \\
	\hline
	Epileptic seizure detection using posterior probability-based convolutional neural network classifier \cite{sivasankari_epileptic_2024}  & Hilberg-Huang transform; posterior probability-based CNN & Bonn \\
	\hline
	\hline
	Integrative Approach for Epileptic Seizure Detection: A Comparative Analysis \cite{sharmila_integrative_2024} & CNN, X-Trees & CHB-MIT \\
	\hline
	Enhanced Epileptic Seizure Detection Through Graph Spectral Analysis of EEG Signals \cite{sharma_enhanced_2024} & Graph-based Fourier transform & Bonn, NSC-ND \\
	\hline
	Machine Learning-Based Feature Extraction Techniques for Epileptic Seizure Detection Using EEG Bio-signals \cite{sharma_machine_2024} & Random Forest (RF), Naive Bayes, KNN, XGBoost, etc & CHB-MIT \\
	\hline
	Exploring Supervised Machine Learning Classifiers for Epileptic Seizure Detection over Two Distinct Preprocessed Datasets \cite{sharma_exploring_2024} & RF, SVM, KNN, XGBoost & CHB-MIT, SNMC \\
	\hline
	Epileptic seizure prediction based on Convolutional neural networks and optimization techniques \cite{shanmugam_epileptic_2024} & CNN + XGBoost & CHB-MIT \\
	\hline
	Attention-based Deep Learning for Epileptic Seizure Type Detection \cite{shankar_attention-based_2024} & LSTM + self-attention & TUH  \\
	\hline
	Epileptic seizure detection using scalogram-based hybrid CNN model on EEG signals \cite{sadam_epileptic_2024} & CNN + SVM & CHB-MIT \\
	\hline
	Epileptic Seizure Detection using DWT based on MRMR Feature Selection Method \cite{rizki_epileptic_2024} & wavelet transform + miminum redudnancy maximum relevance feature selection + SVM & Bonn  \\
	\hline
	A hybrid 1D CNN-BiLSTM model for epileptic seizure detection using multichannel EEG feature fusion \cite{ravi_hybrid_2024} & 1D CNN + BiLSTM & CHB-MIT \\
	\hline
	A compact spatial attention model for automated epileptic seizure detection using multichannel EEG \cite{ravi_compact_2024} & CNN + attention module & CHB-MIT  \\
	\hline
	Anchoring temporal convolutional networks for epileptic seizure prediction \cite{rao_anchoring_2024} & sample entropy + dilated causal convolutional NNs & Freiburg, CHB-MIT \\
	\hline
	Effective Epileptic Seizure Detection Using Enhanced Salp Swarm Algorithm-based Long Short-Term Memory Network \cite{rani_effective_2024} & enhanced feature selection + LSTM & Bonn, TUH, Bern-Barcelona \\
	\hline
	IoT and cloud computing-based automated epileptic seizure detection using optimized Siamese convolutional sparse autoencoder network \cite{ramkumar_iot_2024} & Q-wavelet transform + Siamese convolutional sparse autoencoder & Bonn, TUH \\
	\hline
	Epileptic Seizure Prediction Based on Synchroextracting Transform and Sparse Representation \cite{ra_epileptic_2024} & Synchroextracting Transform + Sparce Representation + KNN & CHB-MIT, Bonn \\
	\hline
	Epileptic Seizure Prediction Using Stacked CNN-BiLSTM: A Novel Approach \cite{quadri_epileptic_2024} & 1D CNN + BiLSTM & CHB-MIT \\
	\hline
	Semi-Supervised Seizure Prediction Based on Deep Pairwise Representation Alignment of Epileptic EEG Signals \cite{qi_semi-supervised_2024} & data augmentations + semi-supervised seizure prediction & CHB-MIT, AES-SPC (KEPC) \\
	\hline
	Efficient Approach for Epileptic Seizure Classification and Detection based on Genetic Algorithm with CNN-RNN Classifier \cite{priyanka_efficient_2024} & improved genetic algorithm + CNN-LSTM-GRU & Bonn \\
	\hline
	A Novel Framework for Epileptic Seizure Detection Using Electroencephalogram Signals Based on the Bat Feature Selection Algorithm \cite{pouryosef_novel_2024} & DWT + BAT feature selection + Naive Bayes or KNN / etc & Bonn \\
	\hline
	Prediction of epileptic seizure based on image information derived from focal electrodes and deep classifiers \cite{pires_prediction_2024} & various methods & EPILEPSIAE \\
	\hline
	Enhanced Epileptic Seizure Detection Based on Information Fusion Techniques \cite{pedram_enhanced_2024} & RF, KNN, SVM, decision tree & Bonn \\
	\hline
	A Robust Machine Learning Method for Real-Time Epileptic Seizure Detection in EEG \cite{pawar_robust_2024} & logistic regression & CHB-MIT \\
	\hline
	Utilizing Pretrained Vision Transformers and Large Language Models for Epileptic Seizure Prediction \cite{parani_utilizing_2024} \textdagger & Vision Transformers, LongFormer LLM & TUSZ (part of TUH) \\
	\hline
	Empirical Mode Decomposition for Deep Learning-Based Epileptic Seizure Detection in Few-Shot Scenario \cite{pan_empirical_2024} &  Empirical Mode Decomposition + DFT + PSD + CNN & Bonn \\
	\hline
	New approaches to epileptic seizure prediction based on EEG signals using hybrid CNNs \cite{nour_new_2024} & 1D CNN & Bonn \\
	\hline
	Detection of Epileptic Seizure from EEG Signals Using Majority Rule Based Local Binary Pattern \cite{nithya_detection_2024} & majority rule-based feature generation + KNN or SVM & Bonn, Freiburg \\
	\hline
	Epileptic Seizure Prediction From EEG Signal Recording Using Energy and Dispersion Entropy with SVM Classifier \cite{nabila_epileptic_2024} & DWT + SVM & CHB-MIT \\
	\hline
	Epileptic Seizure Detection Using Energy Thresholding \cite{mourad_epileptic_2024} & time-domain energy thresholding & CHB-MIT \\
	\hline
	Hybridization of Sparrow Search Algorithm and Sine Cosine Algorithm for Epileptic Seizure Detection \cite{mohapatra_hybridization_2024} & DWT + Adaptive Neuro Fuzzy Inference System & CHB-MIT \\
	\hline
	Real-Time Epileptic Seizure Prediction Method With Spatio-Temporal Information Transfer Learning \cite{meng_real-time_2024} & RNN + transfer learning & CHB-MIT, Siena \\ 
	\hline
	Epileptic Seizure Detection Using Quantum Support Vector Classifier \cite{meesala_epileptic_2024} & quantum SVM & Bonn \\
	\hline
	Epileptic Seizure Detection using Variational Quantum Classifier \cite{meesala_epileptic_2024-1} & Variation Quantum Classifier & Bonn \\
	\hline
	Explainable Graph Neural Networks for EEG Classification and Seizure Detection in Epileptic Patients \cite{mazurek_explainable_2024} \textdagger & GNN + attention layers + feature \& graph explanations & CHB-MIT \\
	\hline
	SH-OSP: A Hybrid Algorithm Using Spotted Hyena Optimizer Enabled with Optimal Stochastic Process for Epileptic Seizure Detection \cite{mallik_sh-osp_2024} & DWT + ANOVA + SVM + advanced optimization. Claimed SoTA & CHB-MIT \\
	\hline
	Parallel Dual-Branch Fusion Network for Epileptic Seizure Prediction \cite{ma_parallel_2024} & MFCC + CNN + Transformer & CHB-MIT \\
	\hline
	MTL-SSU: A Multi-Task Self-Supervised Learning Framework for Epileptic Seizure Prediction \cite{lou_mtl-ssu_2024} & U-Net + linear classifier & CHB-MIT \\
	\hline
	Forecasting events in multidimensional electroencephalographic brain data: Application to epileptic seizure prediction \cite{liu_forecasting_2024} & Neural mass model + LSTM & Custom (St. Vincent's Hospital, Melbourne) \\
	\hline
	Epileptic seizure prediction based on EEG using pseudo-three-dimensional CNN \cite{liu_epileptic_2024} & pseudo-3D CNN + biConvLSTM-3D & CHB-MIT \\
	\hline
	Multi-dimensional hybrid bilinear CNN-LSTM models for epileptic seizure detection and prediction using EEG signals \cite{liu_multi-dimensional_2024} & self-attention + bilinear NN & CHB-MIT, AES-SPC \\
	\hline
	Epileptic Seizure Detection in SEEG Signals Using a Unified Multi-Scale Temporal-Spatial-Spectral Transformer Model \cite{li_epileptic_2024} \textdagger & channel embedding temporal-spatial-spectral transformer & Custom (LTSZ), CHB-MIT, Bonn \\ 
	\hline
	Epileptic Seizure Detection Using a Lightweight Network based on Style-Controlling Normalization \cite{li_epileptic_2024-1} & custom NN architecture (Faster-NormNet) & Bonn, custom \\
	\hline
	Detection of epileptic seizure in EEG signals using machine learning and deep learning techniques \cite{kunekar_detection_2024} & LSTM & Bonn \\
	\hline
	Preictal period optimization for deep learning-based epileptic seizure prediction \cite{koutsouvelis_preictal_2024} & CNN + Transformer & CHB-MIT  \\
	\hline
	A Mutual Information-Based Many-Objective Optimization Method for EEG Channel Selection in the Epileptic Seizure Prediction Task \cite{kouka_mutual_2024} & ConvLSTM + particle swarm optimization & CHB-MIT \\
	\hline
	Epileptic Seizure Detection in EEG Signals Using Machine Learning and Deep Learning Techniques \cite{kode_epileptic_2024} & XGBoost, 1D-CNN, etc. & Bonn \\
	\hline
	Improved Feature Space for EEG-based Epileptic Seizure Detection Using Signal Processing Techniques \cite{kiran_improved_2024} &DWT + improved feature space + SVM & Bonn \\
	\hline
	Explainable AI for epileptic seizure detection in Internet of Medical Things \cite{khan_explainable_2024} & LIME, SHAP & CHB-MIT \\
	\hline
	Feature Fusion for Epileptic Seizure Detection from EEG Data: A CNN- Based Approach with High Precision \cite{kavya_feature_2024} & spectrograms + CNN & Bonn, NSC-ND \\
	\hline
	EEG Conformer Model Based Epileptic Seizure Prediction Using Deep Learning \cite{kasthuri_eeg_2024} & convolutions + Transformer & CHB-MIT \\
	\hline
	An improved GBSO-TAENN-based EEG signal classification model for epileptic seizure detection \cite{kantipudi_improved_2024} & advanced feature selection and optimization + temporal activation expansive NN & Bonn, CHB-MIT \\
	\hline
	Integrated TSVM-TSK fusion for enhanced EEG-based epileptic seizure detection: Robust classifier with competitive learning \cite{kalpana_integrated_2024} & twin SVM + fuzzy classifier & Bonn \\
	\hline
	EpiNET: An Optimized, Resource Efficient Deep GRU-LSTM Network for Epileptic Seizure Prediction \cite{kalita_epinet_2024} & GRU-LSTM & CHB-MIT \\
	\hline
	Epileptic Seizure Detection using Denoising Autoencoder \cite{joshi_epileptic_2024} & denoising autoencoder + SVM / RF / CNN & CHB-MIT, Bonn \\
	\hline
	Sequential graph convolutional network and DeepRNN based hybrid framework for epileptic seizure detection from EEG signal \cite{jibon_sequential_2024} & Sequential Graph Convolution Network + Deep Recurrent NN + GRU & CHB-MIT, TUH \\
	\hline
	Epileptic Seizure Prediction Using Spatiotemporal Feature Fusion on EEG \cite{ji_epileptic_2024} & Graph Attention + Temporal Convolutions & CHB-MIT \\
	\hline
	Domain adaptation for EEG-based, cross-subject epileptic seizure prediction \cite{jemal_domain_2024} & domain adaptation networks & CHB-MIT, Siena \\
	\hline
	Adaptive Bi-LSTM-based Epileptic Seizure Prediction from EEG Signals Using Deep Learning Algorithm \cite{jamunadevi_adaptive_2024} & Bi-LSTM & Bonn, AES-SPC \\
	\hline
	Support Vector Machine-Based Epileptic Seizure Detection Using EEG Signals \cite{himalyan_support_2024} & DWT + SVM & CHB-MIT (?) \\
	\hline
	A Multi Representation Deep Learning Approach for Epileptic Seizure Detection \cite{hermawan_multi_2024} & FFT or STFT + CNN-LSTM & AES-SPC, Melbourne \\
	\hline
	Imbalance-aware Machine Learning for Epileptic Seizure Detection \cite{henni_imbalance-aware_2024} & graph-based feature selection + synthetic minority oversampling & TUSZ (part of TUH) \\
	\hline
	Epileptic Seizure Prediction from EEG Signals using Machine Learning \cite{helgesen_epileptic_2024} &DWT + KNN, RF, etc. & CHB-MIT \\
	\hline
	SeizureLSTM: An optimal attention-based trans-LSTM network for epileptic seizure detection using optimal weighted feature integration \cite{he_seizurelstm_2024} & Q Wavelet transform + 1D-CNN + LSTM & CHB-MIT \\
	\hline
	A lightweight 1D-CNN-GRU model for epileptic seizure prediction \cite{he_lightweight_2024} & 1D-CNN-GRU & CHB-MIT \\
	\hline
	NeuroWave-Net: Enhancing epileptic seizure detection from EEG brain signals via advanced convolutional and long short-term memory networks \cite{hassan_neurowave-net_2024} & CNN+LSTM & Bonn \\
	\hline
	Riemannian Manifold-based Epileptic Seizure Detection Using Transfer Learning and Artifact Rejection Techniques \cite{hassan_riemannian_2024} \textdagger & anomaly detection; Riemann mainfold-based, potato-based features; other methods & CHB-MIT, TUSZ (part of TUH) \\
	\hline
	Advanced Machine Learning Techniques for Precise EEG Analysis and Epileptic Seizure Detection \cite{hasan_advanced_2024} & RF & Bonn \\
	\hline
	Analysis and Design of a Compute-in-Memory System for Epileptic Seizure Detection System \cite{gulfaraz_analysis_2024} & fixed-point NN, compute-in-memory NN & CHB-MIT \\
	\hline
	Two-Stage Approach With Combination of Outlier Detection Method and Deep Learning Enhances Automatic Epileptic Seizure Detection \cite{grubov_two-stage_2024} & wavelets + SVM-based outlier detection + CNN & Custom (83 subjects) \\
	\hline
	Error-aware CNN improves automatic epileptic seizure detection \cite{grubov_error-aware_2024} & wavelets + ResNet-18 & Custom \\
	\hline
	Enhanced Detection of Epileptic Seizure Using Supervised and Unsupervised Algorithms \cite{gopalakrishnan_enhanced_2024} & DWT + K-means, SVM, RF, KNN & AES-SPC (?)  \\
	\hline
	The utility of Multicentre Epilepsy Lesion Detection (MELD) algorithm in identifying epileptic activity and predicting seizure freedom in MRI lesion-negative paediatric patients \cite{goel_utility_2024} & adapted MLD algorithm (using MRI + EEG data) for focal epilepsy detection & Custom (SEEG, MRI, PET; 28 children)  \\
	\hline
	Quantification of EEG Characteristics for Epileptic Seizure Detection and Monitoring of Anaesthesia Using Spectral Analysis \cite{ghuli_quantification_2024} & power spectral analysis & public data from MATLAB\textregistered File Exchange \\
	\hline
	Online Epileptic Seizure Detection in Long-term iEEG Recordings Using Mixed-signal Neuromorphic Circuits \cite{gallou_online_2024} & Spiking NN & SWEC-ETHZ \\
	\hline
	Epileptic Seizure Detection Using a Recurrent Neural Network With Temporal Features Derived From a Scale Mixture EEG Model \cite{furui_epileptic_2024} &Feature extraction based on stochastic modeling + RNN & Custom \\
	\hline
	Enhanced Epileptic Seizure Detection through Wavelet-Based Analysis of EEG Signal Processing \cite{fredes_enhanced_2024} & DWT + SVM & Neonate, CHB-MIT \\
	\hline
	Enhanced Epileptic Seizure Detection: Convolution Neural Net and Features Selection in EEG Signals \cite{fatma_enhanced_2024} & CNN & CHB-MIT \\
	\hline
	EpiNet: A Hybrid Machine Learning Model for Epileptic Seizure Prediction using EEG Signals from a 500 Patient Dataset \cite{esha_epinet_2024} & SVM, XGBoost, etc & Bonn \\
	\hline
	Hybrid Stacking Model for Automatic Epileptic Seizure Detection Using Electroencephalogram Signals \cite{egorova_hybrid_2024} & DT, RF, SVM & TUH \\
	\hline
	Seizure stage detection of epileptic seizure using convolutional neural networks \cite{dutta_seizure_2024} & CNN & Freiburg, CHB-MIT, TUH \\
	\hline
	Hyperdimensional Computing With Multiscale Local Binary Patterns for Scalp EEG-Based Epileptic Seizure Detection \cite{du_hyperdimensional_2024} & Multiscale local binary patterns-based hyperdimensional computing & Beirut \\
	\hline
	Epileptic Seizure Detection with an End-to-End Temporal Convolutional Network and Bidirectional Long Short-Term Memory Model \cite{dong_epileptic_2024} & Temporal CN +BiLSTM & CHB-MIT \\
	\hline
	Integrative Approach for Automated Epileptic Seizure Detection: EEG-MRI Hybrid Model Utilizing Deep Learning Techniques \cite{dharani_integrative_2024} & RNN + CNN + fusion model & Warsaw EEG + MRI \\
	\hline
	A Novel Dual-Branch Network for Comprehensive Spatiotemporal Information Integration for Eeg-Based Epileptic Seizure Detection \cite{deng_novel_2024} &Transformer + CNN + GCN for channel graph & CHB-MIT, Siena \\
	\hline
	Time, Frequency, and Time-Frequency Feature Sets for Deep Learning Methods in Epileptic Seizure Prediction \cite{dasdemir_time_2024} & various feature extraction methods & CHB-MIT \\
	\hline
	Epileptic Seizure Detection from Decomposed EEG Signal through 1D and 2D Feature Representation and Convolutional Neural Network \cite{das_epileptic_2024} \textdagger & empirical mode decomposition + CNN & CHB-MIT, PHK (PHK introduced here) \\
	\hline
	A Lightweight Convolutional Neural Network-Reformer Model for Efficient Epileptic Seizure Detection \cite{cui_lightweight_2024} & DWT + CNN-Reformer & CHB-MIT, custom (SH-SDU) \\
	\hline
	SaE-GBLS: an effective self-adaptive evolutionary optimized graph-broad model for EEG-based automatic epileptic seizure detection \cite{cheng_sae-gbls_2024} & self-adaptive evolutionary graph-regularized broad learning system & CHB-MIT, Bonn, AES-SPC, custom \\
	\hline
	An efficient epileptic seizure detection by classifying focal and non-focal EEG signals using optimized deep dual adaptive CNN-HMM classifier \cite{chavan_efficient_2024} & CNN + HMM + Position Attention + Channel Attention & CHB-MIT, Siena, Bern-Barcelona \\
	\hline
	A Hybrid Study for Epileptic Seizure Detection Based on Deep Learning using EEG Data \cite{buldu_hybrid_2024} & CWT or STFT + CNN & Bonn, NSC-ND \\
	\hline
	Enhancing Deep Learning-Based Epileptic Seizure Detection with Generative AI Techniques \cite{bouallagui_enhancing_2024} & CWT + BiLSTM + VAE & CHB-MIT \\
	\hline
	ResneXt-Lenet: A hybrid deep learning for epileptic seizure prediction \cite{borhade_resnext-lenet_2024} & ResneXt + LeNet (CNNs) & CHB-MIT \\
	\hline
	HyEpiSeiD: a hybrid convolutional neural network and gated recurrent unit model for epileptic seizure detection from electroencephalogram signals \cite{bhadra_hyepiseid_2024} \textdagger & 1D-CNN + GRU & Bonn, Khati Renuka  \\
	\hline
	EEG Signals Classification for Epileptic Seizure Detection \cite{bettayeb_eeg_2024} & CNN + BiLSTM & Bonn \\
	\hline
	Hybrid approach for the detection of epileptic seizure using electroencephalography input \cite{basha_hybrid_2024} & GRU + SVM & Bonn \\
	\hline
	A New Epileptic Seizure Prediction Framework Based on Electroencephalography Signals \cite{assim_new_2024} & LSTM & Neonate, CHB-MIT \\
	\hline
	Epileptic Seizure Detection from Eeg Signals with Recurrent Neural Networks Based Classification Model \cite{aslan_epileptic_2024} & wavelets + RNN & Bern-Barcelona \\
	\hline
	Automatic epileptic seizure detection using MSA-DCNN and LSTM techniques with EEG signals \cite{anita_automatic_2024} & Multi-Scale Atrous-based DCNN + LSTM & CHB-MIT; non-epileptic \\
	\hline
	Epileptic Seizure Detection and Analysis Using Machine Learning \cite{aniruddha_prabhu_bs_epileptic_2024} & LightGBM & Bonn \\
	\hline
	Epileptic Seizure Prediction on EEG Data using a Firefly Algorithm trained with Deep Neural Networks \cite{anandan_epileptic_2024} & Firefly algorithm + GRU & TUH \\
	\hline
	LMPSeizNet: A Lightweight Multiscale Pyramid Convolutional Neural Network for Epileptic Seizure Detection on EEG Brain Signals \cite{alsaadan_lmpseiznet_2024} & multiscale CNN & CHB-MIT \\
	\hline
	EEG-Based Patient Independent Epileptic Seizure Detection Using GCN-BRF \cite{alqirshi_eeg-based_2024} & RF-based feature selection + GCN & CHB-MIT \\
	\hline
	Epileptic Seizure Prediction Using One Channel EEG Signal and 2 D-Convolutional Neural Networks \cite{alizadeh_epileptic_2024} & CNN & CHB-MIT \\
	\hline
	Energy efficient FPGA implementation of an epileptic seizure detection system using a QDA classifier & Quadratic Discriminant Analysis & Bonn \\
	\hline
	Multi-Feature Fusion-Based Convolutional Neural Networks for EEG Epileptic Seizure Prediction in Consumer Internet of Things \cite{ahmad_multi-feature_2024} & CNN + BiLSTM & CHB-MIT  \\
	\hline
	An efficient feature selection and explainable classification method for EEG-based epileptic seizure detection \cite{ahmad_efficient_2024} & DWT + tree-based bagging + SHAP & Bonn \\ 
	\hline
	Robust Epileptic Seizure Detection Based on Biomedical Signals Using an Advanced Multi-View Deep Feature Learning Approach \cite{ahmad_robust_2024} & 1D-CNN + multi-view forest + SHAP & Bonn \\
	\hline
	Epileptic seizure detection and classification of EEG signal using k-nearest neighbors (KNN) compared with ANFIS-adaptive network-based fuzzy inference system \cite{adusumilli_epileptic_2024} & KNN & Bonn \\
	\hline
	Detection of epileptic seizure using EEG signals analysis based on deep learning techniques \cite{abdulwahhab_detection_2024} & CNN + RNN + LSTM & CHB-MIT, Bonn \\
	\hline
	A Novel Unsupervised approach for accurate epileptic seizure detection \cite{abdallah_novel_2024} & Deep Gaussian Mixture Model & CHB-MIT \\
	\hline
	Cross-site generalization using attention layer for epileptic seizure detection \cite{abdallah_cross-site_2024} & 1D-CNN + LSTM + attention & CHB-MIT, custom \\
	\bottomrule
\end{longtable}
\end{center}

\begin{center}
	\begin{longtable}{p{12.5cm}|c}
    \caption{Review papers on epileptic seizure detection published in 2024 and January 2025} \label{table:reviewPapers} \\
		\toprule
		Paper Title & Citations \\
		\midrule\midrule
		A Comprehensive Analysis of ML and DL Approaches for Epileptic Seizure Prediction \cite{yalabaka_comprehensive_2024} & 0 \\
		\hline
		EEG-based epileptic seizure detection using deep learning techniques: A survey \cite{xu_eeg-based_2024}  & 1 \\
		\hline
		Research progress of epileptic seizure prediction methods based on EEG \cite{wang_research_2024} & 2 \\
		\hline
		Epileptic Seizure Prediction through ML And DL Models: A Survey \cite{viswanath_epileptic_2024} & 0 \\
		\hline
		Comparative Analysis of Epileptic Seizure Detection Techniques \cite{velvizhi_comparative_2024} & 0 \\
		\hline
		Machine Learning Techniques in Epileptic Seizure Detection: A Comprehensive Review \cite{thakare_machine_2024} & 0 \\
		\hline
		A systematic review of cross-patient approaches for EEG epileptic seizure prediction \cite{shafiezadeh_systematic_2024} & 0 \\
		\hline
		EEG Channel Selection for Epileptic Seizure Prediction \cite{marinis_eeg_2024} & 1 \\
		\hline
		Prediction of Epileptic Seizure from EEG Signal by DWT And ANN Technique-A Review \cite{khandekar_prediction_2024} & 0 \\
		\hline
		Extensive Review of Epileptic Seizure Detection Techniques: Performance, Achievements, and Future Directions \cite{fatma_extensive_2024} & 0 \\
		\hline
		Analysis of Epileptic Seizure Detection Using Deep Learning Algorithms \cite{devi_analysis_2024} & 0 \\
		\hline
		Review of Machine and Deep Learning Techniques in Epileptic Seizure Detection using Physiological Signals and Sentiment Analysis \cite{dash_review_2024} & 35 \\
		\hline
		Validation of artificial intelligence for epileptic seizure detection: Moving from research into the clinic \cite{dan_validation_2024} & 1 \\
		\hline
		Comparison between epileptic seizure prediction and forecasting based on machine learning \cite{costa_comparison_2024} & 8 \\
		\hline
		A systematic review of deep learning algorithms utilising electroencephalography signals for epileptic seizure detection \cite{choudhary_systematic_2024} & 0 \\
		\hline
		Unleashing the potential of spiking neural networks for epileptic seizure detection: A comprehensive review \cite{cherian_unleashing_2024} & 2 \\
		\hline
		EEG Signal Classification for Epileptic Seizure Detection: A Review of Machine Learning Approaches \cite{arora_eeg_2024} & 0 \\
		\bottomrule
	\end{longtable}
\end{center}

\begin{center}
	\begin{longtable}{p{6cm}|c|p{6cm}}
    \caption{Miscellaneous studies on epileptic seizure detection in 2024 and January 2025} \label{table:miscStudies} \\
		\toprule
		Paper Title & Citations & Main Contribution \\
		\midrule\midrule
		Canine EEG Helps Human: Cross-Species and Cross-Modality Epileptic Seizure Detection via Multi-Space Alignment \cite{wang_canine_2025} & 1 &Multi-space alignment framework based on cross-species, cross-modality EEG by employing DL techniques \\
		\hline
		EEG Opto-Processor: Epileptic Seizure Detection Using Diffractive Photonic Computing Units \cite{yan_eeg_2024} & 2 & High-performance hardware + software\\
		\hline
		RRAM-Based Bio-Inspired Circuits for Mobile Epileptic Correlation Extraction and Seizure Prediction \cite{wang_rram-based_2024} & 0 & Energy-efficient memristor-based circuits for EEG feature extraction and seizure detection\\
		\hline
		MEMS resonator-based reservoir computing for epileptic seizure prediction \cite{kawaguchi_mems_2024} &  0 & Developed a power-effecient microelectromechanical system for seizure prediction \\
		\hline
		Efficient EEG motion artifact elimination framework for ambulatory epileptic seizure detection application \cite{murali_krishn_efficient_2024} &  1 & Motion artifact removal using Singular Spectrum Analysis decomposition and Relative Total Variation filter \\
		\hline
		Epileptic seizure detection with Tiny Machine Learning-a preliminary study \cite{lemoine_epileptic_2024} & 0 & Models for seizure detection using Arduino BLE Sense + Raspberry Pi Zero \\
		\hline
		Benchmarking RRAM Crossbar Arrays for Epileptic Seizure Prediction \cite{khan_benchmarking_2024} &  0 & Assessing RRAM-based algorithms for seizure prediction via RRAM simulation \\
		\hline
		GPS based IoT Module for Vehicle Safety in Epileptic Seizure Detection and Alcohol Monitoring \cite{jagtap_gps_2024} & 0 & Arduino-based device for epileptic seizure detection \\
		\hline
		A Smart IoT-Cloud Framework with Adaptive Deep Learning for Real-Time Epileptic Seizure Detection \cite{hussein_smart_2024} & 1 & Proposed (but not implemented) an IoT infrastucture design for timely seizure detection \\
		\hline
		Cyber Security Based Application-Specific Integrated Circuit for Epileptic Seizure Prediction Using Convolutional Neural Network \cite{dhandayuthapani_cyber_2024} &  0 & Proposed a Field Programmable Gate Array-accelerated CNN  for seizure prediction \\
		\hline
		Epileptic seizure detection using CHB-MIT dataset: The overlooked perspectives \cite{ali_epileptic_2024} & 2 & Proposed a revised approach to assessing the performance of models, more practical than ROC/AUC \\
		\bottomrule
	\end{longtable}
\end{center}

\begin{center}
	\begin{longtable}{p{6cm}|c|p{6cm}}
		\caption{Papers on EEG-based epileptiform activity detection (2024 - Jan 2025) } \label{table:epileptiform} \\
		\toprule
		Paper Title (\textdagger  -- open-source code) & Citations & Main Contribution \\
		\midrule\midrule
		Detection of interictal epileptiform discharges using transformer based deep neural network for patients with self-limited epilepsy with centrotemporal spikes \cite{tong_detection_2025} & 2 & Transformer-based IED detector. Evaulated on a private dataset and TUEV \\
		\hline
		Automatic Multi-label Classification of Interictal Epileptiform Discharges (IED) Detection Based on Scalp EEG and Transformer \cite{rao_automatic_2024} & 0 & CNN + Transformer IED detector. Evaluated on a private dataset \\
		\hline
		Interictal Epileptiform Discharge Detection Using Time-Frequency Analysis and Transfer Learning \cite{munia_interictal_2024} & 0 & ResNet-50 applied to TUEV \\
		\hline
		vEpiNet: A multimodal interictal epileptiform discharge detection method based on video and electroencephalogram data \cite{lin_vepinet_2024} & 3 & Video (body and face detected with YOLOv5) + EEG (processed with EfficientNet2-S) for IED detection. Tested on a private dataset (vEpiSet is an EEG component of this dataset) \\
		\hline
		LightIED: Explainable AI with Light CNN for Interictal Epileptiform Discharge Detection \cite{inoue_lightied_2024} & 0 & CNN for IED detection + explanation with Grad-CAM. Evaluated on a private dataset \\
		\hline
		Detection of Interictal epileptiform discharges with semi-supervised deep learning \cite{de_sousa_detection_2024} & 2 & Semi-supervised VAE for IED detection. Evaluated on a private dataset \\
		\hline
		Robust compression and detection of epileptiform patterns in ECoG using a real-time spiking neural network hardware framework \cite{costa_robust_2024} \textdagger & 16 & IED and high frequency oscillation (HFO) detector on a low power consuming chip. Dataset made publicly available  \\
		\hline
		Advancing epilepsy diagnosis: A meta-analysis of artificial intelligence approaches for interictal epileptiform discharge detection \cite{borges_camargo_diniz_advancing_2024} & 1 & An analysis of 23 studies (IED detection models) \\
		\hline
		Method for cycle detection in sparse, irregularly sampled, long-term neuro-behavioral timeseries: Basis pursuit denoising with polynomial detrending of long-term, inter-ictal epileptiform activity \cite{balzekas_method_2024} \textdagger & 2 & A method for IED rate analysis (IED detection not covered) \\
		\hline
		A review of signal processing and machine learning techniques for interictal epileptiform discharge detection \cite{abdi-sargezeh_review_2024} & 14 & A review of IED detection approaches \\
		\bottomrule
	\end{longtable}
\end{center}

\end{appendices}

\bibliography{sn-bibliography}

\end{document}